# Unmanned Surface Vehicle: Yaw Modeling and Identification

Ahsan Tanveer[*1] and Sarvat Mushtaq Ahmad[1]

[1]Faculty of Mechanical Engineering, GIK Institute of Engineering Sciences & Technology, Swabi, Pakistan

[*]ahsantanveer3883@gmail.com

*Abstract –* In this article, a simplified modeling and system identification procedure for yaw motion of an unmanned surface vehicle (USV) is presented. Two thrusters that allow for both speed and direction control propel the USV. The outputs of the vehicle under inquiry include parameters that define the mobility of the USV in horizontal plane, such as yaw angle and yaw rate. A linear second order model is first developed, and the unknown coefficients are then determined using data from pool trials. Finally, simulations are carried out to verify the model so that it may be used in a later study to implement various control algorithms.

## I. Introduction

Most real-world systems are nonlinear and have inconsistencies in their definition, as well as dynamic parameters that can be affected by outside disturbances. In order to model these systems and determine their physical properties, assumptions are typically applied. A controller is then developed based on the rudimentary mathematical models of the systems. Additionally, building a model for system identification is more challenging than designing a model for control law development.

There is not much work available in the literature on USVs driven by water jets that is aimed towards modeling or identification [1]. However, there have been several studies on the identification of USVs models for orientation control [2] and autonomous surface vehicles (ASVs) models for guidance and ship maneuvering [3], [4], [5], [6]. Nonetheless, for identification, certain experimentation techniques are employed, as seen in [7] and [8]. [9] identifies a nonlinear model for a sailboat.

The first section of this paper will be devoted to the mathematical modeling of a low-cost indigenously built USV. In the next section, we will go through the methods used to determine the model's parameters and then validate it by contrasting it with actual system response.

## II. USV Dynamics

### A. Vehicle Description

The under-discussion vehicle has a central body that houses the electronics, as shown in Fig. 1. An Arduino Nano microprocessor, an L298n motor driver, and an MPU-6050 inertial measurement unit (IMU) make up the electronics. The USV is under-actuated since it has two thrusters, one attached to the port side and one to the starboard. Through a CAT5 cable, a PC on the ground sends power and control signals to the vehicle.

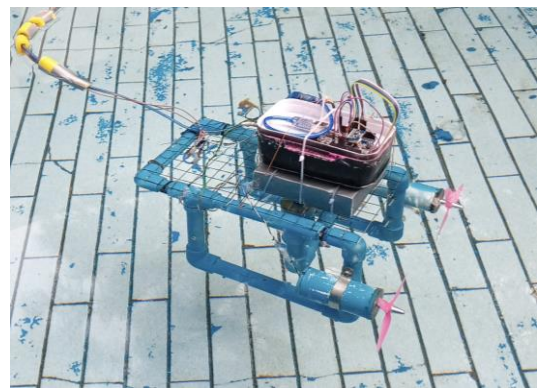

Fig. 1 USV pool trials.

## B. Dynamic Modeling

The following linearized, decoupled equation of motion can be used to represent the yaw mathematical model of a USV with regard to a local body-fixed frame of reference [10].

$$I_z \dot{r} = \tau_{rear} - \tau_d^r \qquad (1)$$

where $I_z$ represents yaw moment of inertia, $r$ yaw rate, $\tau_d^r$ drag in yaw, and $\tau_{rear}$ applied torque of port and starboard-mounted thrusters.

Yaw drag and applied torque is given as [11]:

$$\tau_d^r = b_y r \qquad (2)$$

$$\tau_{rear} = 2la_t u \qquad (3)$$

where $b_y$ represents drag coefficient, $l$ moment arm, $a_t$ thrust coefficient, and $u$ applied input signal in the form of pulse width modulation (PWM).

Using Equation (2) and (3) in Equation (1), we obtain:

$$I_z \dot{r} + b_y r = 2la_t u \qquad (4)$$

Applying Laplace transform to Equation (4) and simplifying:

$$\frac{\psi(s)}{u(s)} = \frac{2la_t}{I_z s^2 + b_y s} \qquad (5)$$

Equation (5) represents yaw model for the USV in frequency domain. Where $r(t) = \dot{\psi}(t)$.

In Equation (1), $b_y$ and $a_t$ need to be determined using data from pool trials. The parameter estimation utilizing system identification is expanded upon in the next section.

## III. SYSTEM IDENTIFICATION FOR PARAMETER ESTIMATION

System Identification (SI) is a dynamic modeling approach that is frequently used in the design of control systems. The three stages involved in identifying a dynamic system model are: model characterization, estimation and validation.

## A. Pool Trials

Experiments are conducted in a controlled pool environment to determine the unknowns in Equation (5). Input-output data is gathered under steady-state conditions during the trials. A square-wave stimulation signal is sent to the USV, and its response is logged, as shown in Fig. 2. The vehicle's output was thus evaluated, and the results show that the data collected provides an accurate approximation of its orientation in the horizonal plane.

## B. Selection of Model Architecture

There is a large selection of linear model structures available. The manner in which each of the different model structures handles the effect of noise is what sets them apart from one another. To ensure agreement with Equation (5), a second order TF model is probed. The final model is:

$$\frac{\psi(s)}{u(s)} = \frac{0.013}{s^2 + 2.08s + 0.46} \qquad (6)$$

Fig. 3 shows the open-loop response of the model. The system response shows that the model is stable.

## C. Validation

The purpose of validation is to evaluate how well the resulting model estimates. A distinct data set is chosen to ensure that the model is not skewed toward certain inputs. Any significant flaws in the obtained model would result in a low fitness score.

Fig. 4 shows the experimental data that was utilized for validation. Model fitness against a distinct dataset is shown in Fig. 5.

A fitness of 59% in the cross-validation test demonstrates the efficacy of the anticipated model. It is therefore maintained that the suggested technique is effective for designing a model-based controller in less time, without requiring considerable mathematical modeling or prior system expertise.

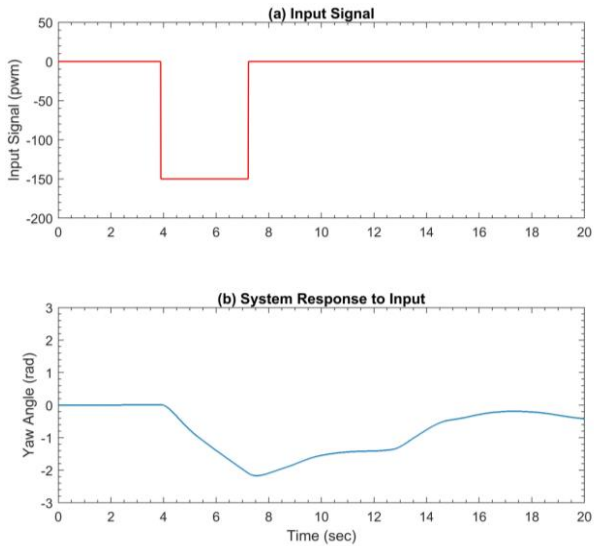

Fig. 2 Training data for model identification.

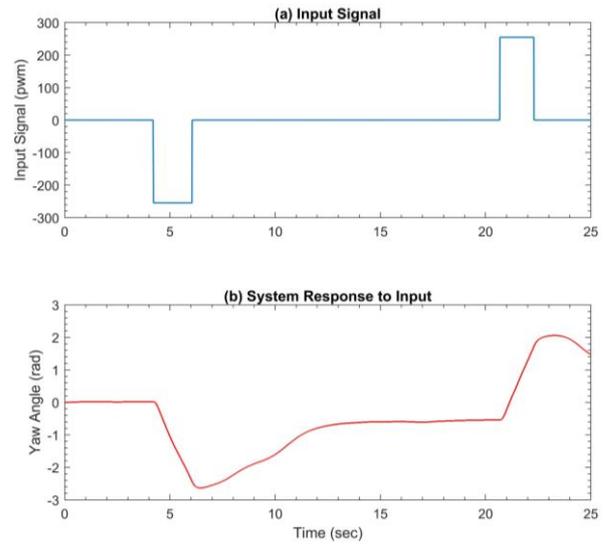

Fig. 4 Testing data for validation.

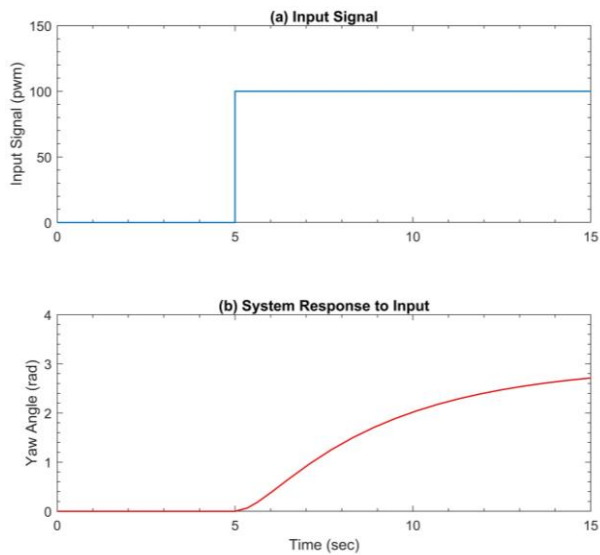

Fig. 3 Yaw model open-loop response.

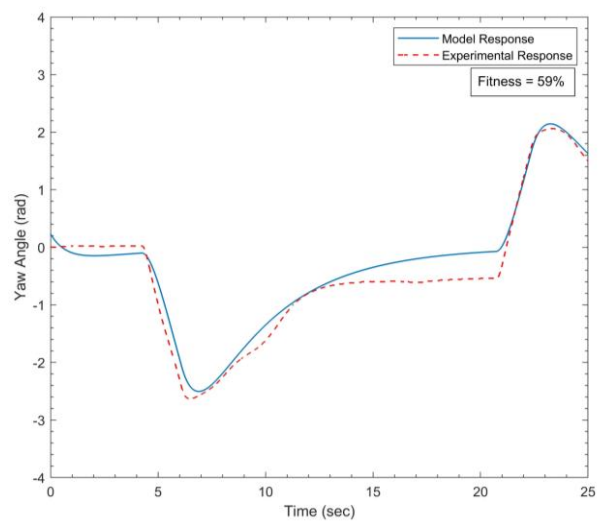

Fig. 5 Cross-validation analysis demonstrating predicted model fitness.

## IV. CONCLUSION

In this paper, an approach for developing a mathematical model for yaw motion of a USV powered by two thrusters is presented. The USV was modeled using the Newton second law of motion. Vehicle parameters were found using data from pool experiments. The suggested method is simple to implement and necessitates only a few tests in a controlled pool setting. The behavior of the resultant identified model is found to be quite comparable to that of the actual vessel.

ACKNOWLEDGMENT

The Ghulam Ishaq Khan Institute of Engineering Science and Technology (GIKI), Swabi, Pakistan, provided funding and other necessary resources for this project.